\documentclass[twocolumn,showpacs,preprintnumbers,amsmath,amssymb,nofootinbib]{revtex4}
\usepackage{graphicx}% Include figurthebibliographye files
\usepackage{dcolumn}
\usepackage{amsmath}          
\usepackage{amssymb}
\usepackage{bm}
\begin{document}
\title{Tsallis meets Boltzmann: $q$-index for a finite ideal gas and its thermodynamic limit}

\author{J. A. S. Lima$^{1}$} \email{jas.lima@iag.usp.br} 
\author{A. Deppman$^{2}$} \email{deppman@if.usp.br}

\affiliation{$^{1}$Departamento de Astronomia (IAG-USP), Universidade de S\~ao Paulo, 05508-090 São Paulo SP,  Brasil}

\affiliation{$^{2}$Instituto de F\'isica, Universidade de S\~ao Paulo, 05508-090 São Paulo, Brasil}

\date{\today}

\begin{abstract}
\noindent Nonadditive Tsallis $q$-statistics has successfully been  applied for a plethora of systems in natural sciences and other branches of knowledge. Nevertheless, its foundations have  been severely criticised by some authors based on the standard additive Boltzmann-Gibbs approach thereby remaining a quite controversial subject. In order to clarify some polemical concepts, the distribution function for an ideal gas with a finite number of point particles and its $q$-index are analytically determined. The two-particle correlation function is also derived. The degree of correlation diminishes continuously with the growth of the number of  particles. The ideal finite gas system is usually correlated, becomes less correlated when the number of particles grows, and is finally, fully uncorrelated  when the molecular chaos regime is reached.  It is also advocated that both approaches can be confronted through a careful kinetic spectroscopic experiment. The analytical results derived here suggest that Tsallis q-statistics may play a  physical role more fundamental than usually discussed in the literature.  
\end{abstract}

\pacs{24.10.Pa; 26.60.+c; 25.75.-q}

\keywords{}

\maketitle
%\vspace {1.0cm}
\noindent{\it 1. Introduction.} The so-called nonextensive $q$-statistics proposed nearly three decades ago by Tsallis \cite{Tsallis1988} has found many applications (from econometry to black hole physics), and prompted studies on the meaning of entropy as a physical and informational concept \cite{2,3,dual}. In certain sense the route followed by Tsallis can be termed a statistical\, ``\textit{top-down}" approach. Actually,  q-statistics  begun with a new nonadditive formula for entropy, a key and abstract concept in the core of kinetic theory and statistical mechanics. Impressive mathematical properties of Tsallis $q$-statistics were  derived based on the ``\textit{top-down}" approach \cite{2,3,dual,EB04,S06}, and the same occurred with the extended physical description of different Hamiltonian systems \cite{2,3,P1,M2,M3,RH06,K2017}.   

As it appears, Boltzmann entropy is just a particular case of Tsallis formula. By defining the q-logarithm and q-exponential functions, $ln_q\, f = \frac{f^{q-1}-1}{q-1}$ and $e_q(f)=[1+(q-1)f]^\frac{1}{q-1}$, so that in the limit $q \rightarrow 1$, $ln_q\,f =ln f$, and $ln_q(e_q(f))=e_q(ln_q(f))=f$, Tsallis entropy in the microcanonical ensemble \cite{Tsallis1988}, provides
%\lim_{q \to 1}\,
immediately the Boltzmann-Gibbs result as a limiting case:
\begin{equation}
S_q = k_B\,ln_q\,W,\,\,\, \lim_{q\rightarrow 1}S_q=k_B\,ln W  \,\,\, (BG)
\end{equation}
where $W$ is the number of microscopic accessible states. The q-entropy is nonadditive in the sense that $S_q(A+B)= S(A) + S(B) + k_B^{-1}(1-q)S(A)S(B)$ which implies the existence of correlations ($q>1$) or anti-correlations ($q<1$) in the system. 

Later on, a \,``\textit{bottom-up}" kinetic approach started by deriving the velocity distribution function (VDF) following the pioneering Maxwell ideas \cite{Max1860}. It was found \cite{SPL1998}  that the one-particle  VDF for an ideal gas assumed in thermodynamic equilibrium can be extended to include $q$-Tsallis version  by relaxing one of the two basic assumptions originally assumed by Maxwell, namely, the factorization of the distribution function: 
\begin{equation}
 f_M(v)d^3v=f(v_x)dv_x \,f(v_y)dv_y \,f(v_z)dv_z \,,
\end{equation}
where $v^2=v_x^2+v_y^2+v_z^2$. When such a working assumption is relaxed by hypothesizing that the velocity distribution is not factorizable, but assuming (like Maxwell) that the velocity space is isotropic, $F_q(v_x,v_y,v_z)=F_q(v^{2})$, the power law velocity $q$-distribution emerging from ensemble approach is obtained \cite{SPL1998} 
\begin{equation}\label{PL} 
 f_q(v)=A_q [1-(q-1) \frac{mv^2}{2k_BT}]^{1/(q-1)}\equiv A_q\,e_q(-\frac{mv^2}{2k_BT})\,
\end{equation}
where $A_q$ is the normalization constant, m is the mass of the particles, $k_B$ is the Boltzmann constant and the limit ${q\rightarrow 1}$ yields $f_q(v) = f_M(v)$. 
%Such an approach suggested that the ideal gas had a nonextensive signature whose origin was not determined. 
A $q$-generalization of the Boltzmann H-theorem was also investigated \cite{LSP2001}. Since the \textit{r.h.s.} of Boltzmann equation is nothing but the total time derivative of the distribution function, its possible modification must appear in the collisional term. In this case, the Boltzmann equation reads:  
%\begin{equation}
%f(v)d^3v=exp_q\left\{\sum_{i=1}^{3} f_i^{q-1}(v_x) ln_q [f(v_i)] dv_x \right\}\,,
%\end{equation}
%where $exp_q(f)=[1-(1-q))f]^{1/(1-q)}$ and $ln_q f=(f^{1-q}-1)/(1-q)$.
\begin{equation}
 \frac{\partial f}{\partial t}  + {\bf v}. \nabla f+ {\bf a}. \frac{\partial f}{\partial {\bf v}}=\left(\frac{\partial f}{\partial t}\right)_{q-coll.} \,, \label{MaxwellTsallis}
\end{equation}
where ${\bf a}$ is the particle acceleration, ${\bf v}$ is the particle velocity and $\left(\frac{\partial f}{\partial t}\right)_{q-coll}$ is the q-collisional term. It was shown that the q-power law (\ref{PL}) is the unique solution under which the proposed q-collisional term is nullified. However, both approaches were unable to determine the $q$-parameter as a function of the extensive quantities. It was constrained using the H-theorem ($q > 0$) but not quantified in terms of the gas properties (for similar studies in the relativistic domain see \cite{PRE05,L02}).

In this \textit{Letter}, beyond to fill this gap we also address other closely related problems/questions underlying Tsallis $q$-statistics: (i) which is the  $q$-index expression for an ideal finite gas far from the molecular chaos regime and how to quantify the  degree of correlation in the Tsallis regime? (ii) which is the nature of such correlations whether ``in or out" of thermodynamic equilibrium? (iii) Is there a ``transition" describing  the emergency of the uncorrelated molecular chaos from the correlated Tsallis regime? Finally, we also discuss the key question: (iv) how Tsallis' regime can experimentally be probed for this unique simple system? 
%As far as we know, all these basic questions remains answered for a  simple system like a finite dilute gas. 
% P

In order to avoid theoretical subtleties and  controversial issues involving the nonadditive entropy definition (see for instance, \cite{dual,Jizha}),  in what follows we adopt the ``\textit{bottom-up}" approach  for answering the points listed above.  Our attention it will be concentrated on the kinetic behavior of an ideal gas with a finite number of particles. This system has been kinetically studied before by several authors, however, focused on the so called thermodynamic limit, $N \rightarrow \infty$ (see for instance \cite{Cerc0,Cerc1, Huang}). This ideal  finite gas is the simplest statistical system for which a ``bottom-up" extended framework can be applied, and, as such, it can be dubbed an ideal $q$-gas in the sense of Tsallis \cite{LS2005}.  

Our leitmotiv is that Tsallis power-law regime and its MB limit must emerge  naturally from the Liouville-kinetic description of finite systems.  As we shall see, for a gas of N particles, the four above aspects  can be elucidated at once.  
%The entropic $q$-index it will be determined as a function of the number of  particles becoming Maxwellian ($q \rightarrow 1$) only in the limit $N \rightarrow \infty$. 
In particular, it will be  demonstrated that the finite ideal gas is usually correlated, becomes less correlated whether the number of particles grows, and, finally, fully uncorrelated  when the molecular chaos regime is reached. 
%This result suggests that nonadditivity can also be present even for moderately correlated systems in thermodynamic equilibrium. 
Based on such results we also discuss a simple spectroscopic experiment to confront Tsallis and Boltzmann approaches, as well as, to delimit when the predicted change of regimes starts in the $q$-gas.

\noindent \textit{2. From Liouville N-particle distribution to the one-particle VDF of a $q$-ideal gas.} In the Liouville description, the state of an ideal gas composed by N particles is a point in the 6N dimensional $\Gamma$-space  whose coordinates are the spatial coordinates and velocities of all particles. The location of the $i^{th}$ particle in the $\Gamma$-space is the 6-dimensional vector $\Sigma_i = (x_{i}, v_{i})$. In addition, the probability that a $\Gamma$-point is found in a unit volume of $\Gamma$-space at time $t$ is given by the N-particle distribution function, $f_{N} (\Sigma_1,....,\Sigma_N$,t). It is normalized 

\begin{equation}\label{NC}
 \int_{-\infty}^{\infty}f_{N}(\Sigma_1,...,\Sigma_N,t) d^6\Sigma_1..... d^6\Sigma_N = 1 
\end{equation}
where $d\Sigma_i=d^{3}x_{i}d^{3}v_{i}$. As widely known, the convective derivative $f_{N}$ satisfies the collisionless Boltzmann equation $df_{N}/dt=0$. Since we are describing velocity independent forces in a time reversible Hamiltonian system, this also means that Tsallis' entropy remains constant. In other words, all the information is in principle available in the Liouville equation. The one and two-particle VDF are the reduced distributions by integrating out all the irrelevant degrees of freedom. 

Let us now consider an homogeneous monatomic gas with particles of mass $m$ in thermal equilibrium at temperature T, within a box of volume $V$. The total energy of the gas reads:
\begin{equation}
E=\frac{m}{2}\sum_{i=1}^{N}v_{i}^{2} = Nm\epsilon \rightarrow \sum_{i=1}^{N}v_{i}^{2} = 2N\epsilon
\end{equation}
where $\epsilon$ is the average energy per unit mass of the gas which is assumed to be a measurable quantity, that is, macroscopically determined. 

The N-particle distribution function for an ideal gas depends only on the kinetic energy. Then, it is natural to assume $f_N$ proportional to a $\delta$-function \cite{Cerc0,Cerc1}:  
\begin{equation}
f_N (\Sigma_1,...,\Sigma_N) = B_N \delta\left(\sum_{i=1}^{N}v_{i}^{2} - 2N\epsilon\right)
\end{equation}
where the constant $B_N$ can be determined from ($\ref{NC}$).  By integrating over all space variables one finds:

\begin{equation}\label{NC1}
 B_N V^{N}\int_{-\infty}^{\infty}\delta\left(\sum_{i=1}^{N}v_{i}^{2} - 2N\epsilon\right)d^{3}v_1... d^3v_N = 1 
\end{equation}
which can be solved transforming to polar coordinates: $\sum_{i=1}^{N}v_{i}^{2}=R^{2}$ with $d^3v_1...d^3v_N=R^{3N-1}dRd^{3N}S$. Note that $d^{3N}S$ is the surface element of a unit sphere in 3N dimensions. As one may check, the normalized $f_N$ now takes the form:

\begin{equation}\label{NC2}
 f_N(v_1,...,v_N) = \frac{2}{\Omega_{3N}{(2N\epsilon)}^{3N-2}V^{N}}\delta\left(\sum_{i=1}^{N}v_{i}^{2} - 2N\epsilon\right)
\end{equation}
where
\begin{equation}\label{SN}
\Omega_{3N}=\int d^{3N}S  = \frac{2\pi^{3N/2}}{\Gamma(3N/2)}
\end{equation}
is the surface of a unit sphere in 3N dimensions \cite{Huang}. 

The one-particle VDF yields the probability to find at random a molecule between $\Sigma_1$ and $\Sigma_1 + d\Sigma_1$ regardless of the $\Gamma$-space position of the other molecules. Now, by taking $v_1=v$ and following the tradition (see for instance \cite{Huang}), we also normalize $f(v)$ to the particle concentration $n=N/V$ 

\begin{equation}\label{D1}
 f(v)=Nf_1=NV^{N-1}\int_{-\infty}^{\infty}f_{N}(v,v_2,...,v_N) d^3v_2...d^3v_N  
\end{equation}
where the term $V^{N-1}$ is nothing but the  volume integrated over the remaining coordinates. By inserting the value of $f_N$ from ($\ref{NC2}$) and using again polar coordinates, the integration over the remaining velocities results:
\begin{equation}\label{D1a}
 f(v)= \frac{n}{(2N\epsilon)^{3/2}}\frac{\Omega_{(3N-3)}}{\Omega_{3N}}\left[1- \frac{v^2}{2N\epsilon}\right]^{(3N-5)/2} 
\end{equation}
and by inserting the $\Omega$-prefactors defined by ($\ref{SN}$), it follows that:

\begin{equation}\label{D2}
 f(v)= n\left(\frac{3}{4\pi\epsilon}\right)^{3/2} \frac{(\frac{3N}{2})^{-3/2}\Gamma(\frac{3N}{2})}{\Gamma{(\frac{3N}{2}-\frac{3}{2})}}\left[1- \frac{v^2}{2N\epsilon}\right]^{(3N-5)/2}. 
\end{equation}
Note that for a finite value of N, this result describes a power-law with \textit{cut-off} in the velocity space,  $v \leq v_m = (2N\epsilon)^{1/2}$. It is readily checked that in the limit $N\rightarrow \infty$ the above expression reduces to the MB result  
\begin{equation}\label{MDF}
\lim_{N\rightarrow \infty}f(v)=f_M (v)= n\left(\frac{m}{2\pi\,k_B\,T}\right)^{3/2}e^{-\frac{mv^{2}}{2\,k_B\,T}}
\end{equation}
where now we have inserted 
\begin{equation}\label{energy}
\epsilon = \frac{3}{2}\frac{k_B T}{m}   
\end{equation}
as usually demonstrated for the thermodynamic limit. This measurable value is macroscopically determined through the pressure of perfect gases,  $P=nk_BT$.  Its validity have been demonstrated  for a dilute gas when the  molecular chaos is already established. However, since the focus of our article is the finite system whose VDF is described by (\ref{D2}),  it is natural to ask whether Eq. ($\ref{energy}$) still remains valid.  In order to show that, let us consider the kinetic gas pressure as predicted by distribution (\ref{D2}) when the cutoff in the particle velocities is taken into account. In this case, the pressure $P$ is defined by \cite{Huang}
%for f(v) given by ($\ref{D2}$) reads:
% P
\begin{equation}
P = \frac{m}{3}\int_{0}^{v_m} f(v)v^{2}d^3 v=\frac{4\pi m}{3}\int_{0}^{v_m}f(v)v^{4}dv
\end{equation}
and takes the following form for a finite gas:
\begin{equation}
P = \frac{4\pi n m}{3}\left(\frac{3}{4\pi\epsilon}\right)^{3/2}D_N\int_{0}^{v_m}v^{4}\left[1- \frac{v^2}{2N\epsilon}\right]^{(3N-5)/2}dv   
\end{equation}
where $D_N = \frac{(\frac{3N}{2})^{-3/2}\Gamma(\frac{3N}{2})}{\Gamma{(\frac{3N}{2}-\frac{3}{2})}}$. By defining a new variable, $x=v^{2}/2N\epsilon$, the integral is reduced to a complete Beta function \cite{AS}, $B(z,w) = \Gamma(z)\Gamma(w)/\Gamma(z+w)$, with $z =5/2$ and $w = (3N-3)/2$. As a result of the integration process, all terms appearing in $D_N$ cancels out and the pressure becomes  $P = \frac{3}{2}\,nm\epsilon$ (the same Maxwellian value!). This means that  small systems, those with different finite values of ${\bar N}$ and ${\bar V}$, but the same concentration, $n = {\bar N}/{\bar V}=N/V$,  are able to maintain the equilibrium conditions because the mean free path ($\l \simeq 1/n \pi\sigma^{2}$) is not altered ($\sigma$ is the diameter of the hard spheres). Therefore, all these  monatomic gases are endowed with different extensive quantities, but obeys $P=nk_B\,T$. Hence, by equaling both expressions for the pressure, we obtain exactly the same energy per unit mass, $\epsilon$, given by ($\ref{energy}$). 

%these results also clarify some historical aspects related to the calculation of averages and marginal probabilities  of Tsallis $q$-statistics.
%and, have the average collision time, $t_c \sim n\sigma{\bar v}, where $\sigma$ is the cross section and $v$ is average velocity. 

The VDF ($\ref{D2}$) can be rewritten in the standard Tsallis form because the term in the square bracket, $[1-\frac{x^{2}}{\alpha}]^{\alpha}$, is already in the form of a power-law which becomes the MB exponential factor in the limit $\alpha \rightarrow \infty$. To show that, we first observe that  $\Gamma(x)$ is a sharply peaked and rapidly varying function. Thus, by neglecting the factor of a few only in the power-index of ($\ref{D2}$) and comparing with ($\ref{PL}$),  the $q$-parameter is obtained 
\begin{equation}\label{qv}
\frac{3N}{2}= \frac{1}{q-1}\,\,\, \large{\rightarrow}\,\,\,  q= 1 + \frac{2}{3N}
\end{equation}
thereby fixing naturally the MB limit 
\begin{equation}\label{qL}
\lim_{N\rightarrow \infty}\, q(N) = 1.
\end{equation}
%\lim_{x\rightarrow \infty}
Note also that the VDF ($\ref{D2}$) now assumes a Tsallian form (cf.  ($\ref{PL}$)):
\begin{equation}\label{TD}
 f(v)= n\left(\frac{m(q-1)}{2\pi\,k_B\,T}\right)^{3/2}\frac{\Gamma(\frac{1}{q-1})}{\Gamma{(\frac{1}{q-1}-\frac{3}{2})}}\left[1- (q-1)\frac{mv^2}{2k_B\,T}\right]^{\frac{1}{q-1}}.
\end{equation}
As should be expected, the MB velocity distribution is also recovered  from the VDF above by taking the limit $q\rightarrow 1$. 

On the other hand, applying the transformation  $q \rightarrow q' = 1-\frac{2}{3N}$ we find that $q + q'= 2$ which recovers the so-called additive duality relation in this context \cite{dual}. Note also that ($\ref{TD}$) under duality transformation becomes:

\begin{equation}\label{TDD}
 f(v)= n\left(\frac{m(1-q')}{2\pi\,k_B\,T}\right)^{3/2}\frac{\Gamma(\frac{1}{1-q'})}{\Gamma{(\frac{1}{1-q'}-\frac{3}{2})}}\left[1- (1-q')\frac{mv^2}{2k_B\,T}\right]^{\frac{1}{1-q'}}.
\end{equation}
which is also a $q$-power law with cut-off because the duality relation implies that $q'< 1$.

It is also worth noticing that all averages and marginal probabilities were calculated here in the traditional manner. In other words,  $q$-average values of physical quantities and nonstandard calculations of marginal probabilities, as sometimes adopted \cite{dual,QAV,QAV1}, are not necessary in the present context. 

At this point we  recall that several authors  investigated the properties of finite systems based on the\, ``top-down" entropic approach, including different expressions for the $q$-parameter, possible rescaling properties and even conceivable effects for different nonstandard entropies \cite{P1,M2,M3,RH06,K2017}. In principle, the \,``bottom-up" results derived here may also have consequences for such studies. It should be also  stressed that the nonextensive ideal gas is in thermodynamic equilibrium from the very beginning with the standard concepts of temperature, pressure and average energy per particle rigorously preserved. Interestingly, as it will be discussed below, the  homogeneity and simplicity of the $q$-gas distribution facilitate all higher order calculations so that its nonextensivity degree can also be understood based on the analytical two-particle correlation function.

\noindent \textit{3. Correlations from two-particle VDF.} As remarked earlier, the two-particle function $f(v_1,v_2)\equiv f(v,v')$ in this context is a trivial generalization of the one-particle case. It is obtained by integrating out over the irrelevant $N-2$ degrees of freedom. 
%It is usually defined by the probability to find the first particle between $\Sigma_1$ and $\Sigma_1$ + $\Sigma_1$ and the second one
%between  $d\Sigma_1$ and $\Sigma_2$ + $d\Sigma_2$, regardless of the positions of the remaining $N-2$ particles. 
Following standard lines (see also discussion above ($\ref{D1}$)), let us normalize $f(v,v')$ by $N(N-1)\simeq N^{2}$ thereby obtaining: 
%By extending the same approach of $f(v)$, one may show that:  

 \begin{equation*}
f(v,v')=N^{2}V^{N-2}\int_{-\infty}^{\infty}f_{N}(v,v',v_3...,v_N) d^3v_3...d^3v_N 
 \end{equation*}
By inserting $f_N$ from ($\ref{NC2}$),  this integral is solved  as before by changing to polar coordinates. Then, using the same value of $\epsilon$ and also the  $q$-definition from ($\ref{qv}$) we obtain for $N$ greater than a few:
\begin{widetext}
%\begin{equation}
%f(v,v')=n^{2}G_N\left(\frac{m}{2\pi\,k_B\,T}\right)^{3}\left[1-(q-1)\frac{m(v^{2} + %v'^{2})}{2K_BT}\right]^{\frac{1}{q-1}}
% \end{equation}
\begin{equation}\label{CF}
f(v,v')=n^{2}G_q\left(\frac{m}{2\pi\,k_B\,T}\right)^{3}\left[1-(q-1)\frac{m(v^{2} + v'^{2})}{2k_BT}\right]^{\frac{1}{q-1}}\neq f(v).f(v')
\end{equation}
\end{widetext}
where $G_q=(2-q)(3-2q)(4-3q)$. The 2-particle correlation function is also a $q$-power law with cut-off ($q > 1$) satisfying the duality relation. In addition, since the final inequality indicates that it cannot be factorized,  the $q$-gas does not satisfy the extra-mechanical Boltzmann hypothesis, the basis of his H-theorem. However, for very large values of $N$ ($q\rightarrow 1$) and the same limit yields $G_1=1$ thereby obtaining  (compare with  Eq. ($\ref{MDF}$))
%\begin{widetext}
%\begin{eqnarray}
%f(v,v')=n^{2}\left(\frac{m}{2\pi\,k_B\,T}\right)^{3}\left[1-(q-1)\frac{m(v^{2} + %v'^{2})}{2K_BT}\right]^{\frac{1}{q-1}} = f(v).f(v') 
%\end{eqnarray}
%\end{widetext}
\begin{eqnarray}
f(v,v')=n^{2}\left(\frac{m}{2\pi\,k_B\,T}\right)^{3}e^{-\frac{m(v^{2} + v'^{2})}{2k_BT}} = f(v).f(v') 
\end{eqnarray}
showing that the correlation vanishes as expected for the MB limit. These results also suggests that the phenomenological nonextensive Boltzmann H-theorem \cite{LSP2001}, can rigorously be formulated for a $q$-gas.  

The derived index $q=1+2/3N$  together ($\ref{CF}$)  means that the $q$-gas is only moderately correlated. The basic reason is very simple. A finite gas with $10^{2}$ particles departs from the MB limit nearly one part in $10^{-2}$, whereas for a gas with $N=10^{4}$ particles the effect is only one part in $10^{-4}$. Since the system is in thermodynamic equilibrium, such results complete the answer of the problems/questions  (i), (ii), and (iii) outlined in the introduction. Note also that the existence of such correlations much before the molecular chaos regime may be closely related with the energy conservation law during the collisions. In principle, measurements to find these nonextensive effects in a finite ideal gas are possible and of great interest. However, as it will discussed next, experiments with  a relatively small number of particles and a high degree of accuracy are needed. 

\textit{4. Probing nonextensive effects in the $q$-gas.} Let us now discuss how to find possible nonextensive signatures in the finite gas. Several experiments have already verified  Maxwellian \cite{MH90,RD1958} and  non-Maxwellian distributions (see \cite{B96, B99, LSJ2000,B1} and Refs. therein). However, for a finite gas, an additional care must be taken {\it w.r.t.} such experiments: (i) The earlier ones were done when the universal validity of the MB distribution was not being contested, (ii) many experiments involve a very large number of particles (thermodynamic limit), and (iii) power-laws can be easily observed, but, possibly, only for strongly correlated systems. 

On the other hand, predictions related to the $q$-gas  must be tested in its proper regime, that is, far from the MB limit ($N\rightarrow \infty$). For example,  through an experiment where the measured quantity in the MB limit depends only on the temperature whereas in the nonextensive domain $N$ is activated in virtue of the correlations. Some possibilities are related to spectroscopic methods. Probably, the simplest one is the well known  thermal Doppler broadening (TDB) arising from random  motions of the radiating atoms. Its phenomenology is remarkable simple: the light emitted from atoms moving to the observer is blue shifted and those moving away, contribute to a redshift of the spectral lines \cite{RD1958}. For a given atom with velocity $v_z$, the observed wavelength $\lambda$ and the rest frame $\lambda_0$ are related by $\lambda \simeq \lambda_0 (1 + v_z/c)$, where $c$ is the light velocity. Since $v_z$ and $\lambda$ are linearly related, 
%the VDF can directly be transformed to the probability of observing a photon with wavelength $\lambda$. This means that 
the line width must be a q-Gaussian peaked at $\lambda_0$. In the MB limit ($q=1$), the TDB at the rest central wavelength $\lambda_0$ (full width at half maximum) reads \cite{S1986}

\begin{equation}\label{LW1}
\frac{\lambda-\lambda_0}{\lambda_0}\equiv \frac{\Delta \lambda_D}{\lambda_0} = \left(\frac{8k_B\,T}{mc^{2}}\,ln{2}\right)^{1/2}
\end{equation}
whereas for a finite gas, it depends explicitly of $q = 1+\frac{2}{3}N$. By using the q-logarithm definition 
\begin{equation}\label{LW2}
\frac{\Delta \lambda_D}{\lambda_0}= \left(\frac{8k_B\,T}{mc^{2}}\,ln_q\,2\right)^{1/2} \rightarrow  \frac{\Delta \lambda_D}{\lambda_0}=\frac{\Delta \lambda_D}{\lambda_0}(\frac{ln_q\,2}{ln\,2})^{1/2}.
\end{equation}
In terms of $N$ the above Tsallian expression reads:
\begin{equation}\label{LW3}
  \left(\frac{\Delta \lambda_D}{\lambda_0}\right)_T=\left(\frac{\Delta \lambda_D}{\lambda_0}\right)_{MB}\left(\frac{2-2^{1+2/3N}}{3Nln\,2}\right)^{1/2}.
\end{equation}
with the MB result ($\ref{LW1}$) being recovered when $N \rightarrow \infty$. 
%Note  that the thermal broadening for the finite gas is always smaller than the MB result. 
In principle, this kind of experiment may provide a definitive proof of nonextensive effects in this simplest system and the reality of the ideal $q$-gas. 

\textit{5. Final Remarks.} In this article we have discussed the derivation of the velocity distribution function for what we have dubbed the ideal $q$-gas, that is, an ideal gas containing a finite number of particles. 
%Since long ago, it is well known that the limiting case  $N\rightarrow \infty$ recover the MB limit. 
As far as we know, the case for a finite gas and its kinetic connection with  Tsallis statistics has not been investigated before. The main results derived here may be summarized as follow: 
%n the following statements:
%\begin{itemize}

(i) The finite $q$-gas has been discussed by using the conventional kinetic approach without taking the thermodynamic limit. As a result we derived the Tsallis power-law index, $q=1 +\frac{2}{3N}$ (section 2). In the limit ($N\rightarrow \infty$, $q\rightarrow 1$), the standard Maxwell-Boltzmann distribution is recovered. 

(ii) The two-point correlation function of the $q$-gas has also been derived (section 3). It was found that it is only moderately correlated. The existence of correlation seems to be a simple manifestation of the energy conservation law for a relatively small number of particles; a kind of\, ``memory effect" of small systems, vanishing when the molecular chaos regime is attained. The adopted kinetic analysis implies that the entropy is also conserved, and, as such, although correlated, the system is in thermodynamic equilibrium. The $q$-gas provides a simple analytical example that molecular chaos can no longer be considered a central concept for systems with a relatively small number of particles.   

(iii) The properties  of both regimes  (Tsallis and Maxwell-Boltzmann) can be probed preparing systems with different number of particles and volumes, but sharing the same concentration. In principle, several crucial crucial experiments are possible. One of them,  the Doppler broadening of spectral lines, has been discussed with some detail. In Tsallis' regime the prediction of the broadening, ${\Delta \lambda}/{\lambda_0}$, is different from the standard value [section 4, Eqs. ($\ref{LW1}$)-($\ref{LW3}$)], and may reveal the reality of the $q$-gas description.

(iv) The fact that the $q$-gas is only moderately correlated and acted by short-range forces (binary collisions) means that $q$-statistics signatures are not necessarily related to long-range forces or even to strongly correlated systems (including fractals).  Naturally, systems endowed with such ingredients must strength the nonextensive effects thereby explaining the ubiquity of power-laws. However, our results suggests that moderate signatures in less extreme systems, may open a new route to theoretical and experimental understanding of such effects. Hence, although relatively rare in nature, their investigation is by no means less important and challenging.     

Finally, we stress that the approach followed here for the ideal $q$-gas may also be useful to determine the $q$-parameter for similar systems in different fields. If experimentally verified, these results suggest that Tsallis q-statistics for finite systems 
%with a relatively small number of particles 
may play a more fundamental physical role than Boltzmann statistics.
%\end{itemize}

{\bf Acknowledgments:} JASL is partially supported by CNPq (310038/2019-7), CAPES (88881.068485/2014) and FAPESP (LLAMA project, 11/51676-9). AD is partially supported by CNPq (304244/2018-0), INCT-FNA (464898/2014-5) and FAPESP (2016/17612-7). 
%%%%%%%%%%%%%%%%%%%%%%%%%%%%%%%%%%%%%%%%%%
\bibliographystyle{mdpi}

\end{document}